\documentclass[aps,prb,twocolumn,superscriptaddress,showpacs,amsmath,amssymb]
{revtex4-1}
\usepackage{graphicx}
\begin{document}

\title{Monatomic Co, CoO$_2$, and CoO$_3$ Nanowires on Ir(100) and Pt(100) surfaces:
Formation, Structure, and Energetics}

\author{P. Ferstl}
\affiliation{Solid State Physics, Friedrich-Alexander-University
Erlangen-N\"{u}rnberg, D-91058 Erlangen, Germany}

\author{F. Mittendorfer}
\affiliation{ Institut f\"{u}r Angewandte Physik and Center for
Computational Materials Science, Technische Universit\"{a}t Wien,
A-1040 Wien, Austria}

\author{J. Redinger}
\affiliation{ Institut f\"{u}r Angewandte Physik and Center for
Computational Materials Science, Technische Universit\"{a}t Wien,
A-1040 Wien, Austria}

\author{M.A. Schneider} \affiliation{Solid
State Physics, Friedrich-Alexander-University Erlangen-N\"{u}rnberg,
D-91058 Erlangen, Germany}

\author{L. Hammer}
\email{lutz.hammer@fau.de} \affiliation{Solid State Physics,
Friedrich-Alexander-University Erlangen-N\"{u}rnberg, D-91058
Erlangen, Germany}

\begin{abstract}
In this study we investigate the structural and chemical changes of
monatomic CoO$_2$ chains grown self-organized on the Ir(100) surface
[P. Ferstl et al., PRL 117, 2016, 046101] and on Pt(100) under
reducing and oxidizing conditions. By a combination of quantitative
low-energy electron diffraction, scanning tunnelling microscopy, and
density functional theory we show that the cobalt oxide wires are
completely reduced by H$_2$ at temperatures above 320~K and a
3$\times$1 ordered Ir$_2$Co or Pt$_2$Co surface alloy is formed.
Depending on temperature the surface alloy on Ir(100) is either
hydrogen covered ($T<$ 400~K) or clean  and eventually undergoes an
irreversible order-disorder transition at about 570~K. The Pt$_2$Co
surface alloy disorders with the desorption of hydrogen, whereby Co
submerges into subsurface sites. Vice versa, applying stronger
oxidants than O$_2$ such as NO$_2$ leads to the formation of CoO$_3$
chains on Ir(100)  in a 3$ \times$1 superstructure. On Pt(100) such
a CoO$_3$ phase could not be prepared so far, which however, is due
to the UHV conditions of our experiments. As revealed by theory this
phase will become stable in a regime of higher pressure. In general,
the structures can be reversibly switched on both surfaces using the
respective agents O$_2$, NO$_2$ and H$_2$.

\end{abstract}


% 61.05.jh    Low-energy electron diffraction (LEED) and reflection high-energy electron diffraction (RHEED)
% 68.35.bd    metals and alloys
% 68.37.Ef    Scanning tunneling microscopy (including chemistry induced with STM)
% 68.47.Gh    Oxide surfaces
% 82.65.+r    Surface and interface chemistry; heterogeneous catalysis at surfaces

\maketitle

\section{Introduction}
In modern heterogeneous catalysis complex metal-oxide composite
materials are used in numerous processes \cite{ertl1997}. These
so-called ``supported catalysts'' consist of highly dispersed metal
particles on an oxide support. By this, the catalytic activity and
reactivity of both components are combined in a synergetic way. For
example the combination of transition metal oxides (TMOs) and noble
metal nanoparticles is known to exhibit extraordinary performance
for a variety of low temperature catalytic oxidation reactions such
as carbon monoxide conversion\cite{Fu2013,Rodriguez07}. For all
these processes the interface between the metal and the oxide
support plays a crucial role for the catalytic performance
 as it often provides just the active sites for
certain chemical reactions e.g.\ coordinatively unsaturated (CUS)
sites\cite{Fu2013}. Thus optimizing the metal/oxide interface can
maximize the reaction rate.

A disadvantage of supported catalysts is the relatively small
interface area between the metal particles and the oxide.
Conventionally, this can be increased by changing the density of the
nanoparticles \cite{boffa94} or by using more sophisticated
architectures \cite{tandem,Yamada2011}. Another approach towards
high-density oxide-metal interfaces are so-called
\emph{oxide-on-metal inverse catalysts}, which are bimetallic
nanoparticles, where the oxides and thus the active interfaces are
formed right at their surfaces in a preceding activation process or
during the reaction. Such nanostructures in particular with platinum
as the base material find application e.g. as electrocatalysts for
both the anodic oxidation reaction and the cathodic oxygen reduction
reaction in fuel cells \cite{Wu2013}. Moreover, an unprecedented
performance for the oxygen reduction and alcohol oxidation reactions
has been found recently for specially designed Pt-Co nanowires
\cite{Bu2016}. Also, for other Pt-Co nanostructures the formation of
cobalt surface oxides was directly observed during the oxidation in
O$_2$ \cite{Xin2014}.

\begin{figure}[htb]
\centering
\includegraphics[width=\columnwidth]{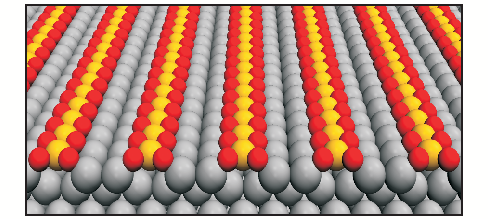}
\caption{Ball model for the 3$\times$1-ordered TMO chain phases
grown on Ir(100) as reported in Ref.\ \onlinecite{Ferstl2016}. (Red:
oxygen; yellow: TM = Mn, Fe, Co, or Ni; gray: Ir substrate.)}
\label{chain_model}
\end{figure}

Despite their technological importance a characterization of these
metal/oxide interfaces on the atomic level is not possible in most
of the cases due to the inaccessibility for standard surface science
techniques. In particular diffraction methods, which would allow for
a quantitative structural analysis, cannot be applied due to the
lack of long range order in these systems. Recently, however, we
found that on the flat Ir(100) surface quasi one-dimensional TMO
chains form self-organized and \textit{strictly periodic}
\cite{Ferstl2016}. This allows to overcome all the mentioned
restrictions and opens direct access to the detailed structure and
composition of the system. A ball model of this chain phase is
displayed in Fig.\ \ref{chain_model}.

Until now, related TMO chain structures could only be prepared via
step decoration of vicinal surfaces \cite{CORh553,MnO2Pd,CoOPd},
where the density and in particular the degree of long range order
is rather limited. In contrast, the high density of one-dimensional
TMO chains leads to a maximum of metal/oxide interfaces for the
given surface area and can therefore be regarded as a model system
for an extremely dispersed case of a bifunctional TMO-noble metal
system.

In the present study we will show using the example of cobalt that
the above mentioned highly ordered chain structures form
self-organized also on the catalytically more relevant Pt(100)
surface. It is rather likely that those TMO chains will appear in
real (Pt-based) catalytically active systems as well making the
presented model studies even more valuable. In the following we also
demonstrate that by interaction with H$_2$, CO, or NO$_2$ these
cobalt oxide nanowires can be reversibly reduced or further oxidized
on both Pt(100) and Ir(100) surfaces, while the one-dimensional
character of the phases is always kept intact. This strict
periodicity of all the different phases allows their structural
characterization on an atomic level not merely by scanning tunneling
microscopy (STM) but also by quantitative low-energy electron
diffraction (LEED) with crystallographic precision. Accompanying
density-functional theory (DFT) calculations do not only corroborate
our structural findings but also give additional insights into the
energetics and stability of the differently oxidized states of the
cobalt chains. The scope of this publication is laid exclusively on
the structural and chemical variations of the chain phases, while
the kinetics and products of the chemical reactions investigated
here will be addressed in forthcoming publications.

\section{Methods}

\subsection{Experimental details and sample preparation}

The LEED and STM experiments were performed in an ultra-high vacuum
chamber consisting of two separately pumped segments. In one part
the sample preparation and the LEED measurements were performed at
an operational pressure in the low 10$^{-10}$~mbar range. Large
series of images of the complete diffraction pattern taken at normal
incidence made up the raw data base for LEED intensity spectra. They
were recorded right after preparation and subsequent cool-down to
about 100~K, which was achieved within 1-5~min depending on the
initial annealing temperature. The time to collect a whole dataset
(45-800~eV in steps of 0.5~eV) took about 10~min only, so that
residual gas contamination can be regarded as negligible. For
further characterization the sample could be transferred to the
second part of the chamber ($p\approx$ 2$\cdot$10$^{-11}$~mbar)
equipped with a room temperature beetle-type STM. Topographs were
recorded with bias voltages from few millivolts to about 1~V applied
to the sample (sample current $\sim$1~nA) using etched tungsten
tips. Further details on the experimental procedures are given in
the Supplemental Material of Ref.~\onlinecite{Ferstl2016}.

The preparation of the Ir(100) and Pt(100) substrates involved an
initial cleaning by ion sputtering (2~keV Ne$^+$, $\sim$10~$\mu$A)
and annealing to 1300~K in an oxygen atmosphere of
$\sim$10$^{-6}$~mbar. The temperature was controlled using a K-type
thermocouple directly attached to the crystal. In order to prepare
the quasi one-dimensional CoO$_2$-chain phase, a 1/3 monolayer (ML)
of cobalt was deposited on either the oxygen covered
Ir(100)-2$\times$1-O phase or the clean, reconstructed Pt(100)
surface using an electron beam evaporator. Subsequently, the surface
was oxidized at an O$_2$ pressure of $\sim$10$^{-7}$~mbar and
annealed to 970~K in case of Ir(100) and to 770~K for Pt(100).

\subsection{LEED calculations}
For the full-dynamical LEED calculations intensity versus energy
curves [``$I(E)$-spectra''] were extracted in an offline evaluation
from the stored stack of LEED images using the programme package
EE2010 \cite{EE2010}, which automatically corrected for the
background of the quasi-elastically scattered electrons. The
post-processing of the spectra involved the normalization to the
measured, energy-dependent primary beam intensity, averaging over
symmetrically equivalent beams, correction for the cosine of the
viewing angle, and slight smoothing. Calculations were performed
using the \textsc{TensErLEED} code \cite{tenserleed} which is based
on the perturbation method Tensor LEED \cite{Heinz95,Rous86}. Due to
the high energy data (up to 800~eV) phase shifts up to $
\ell_{\text{max}}$~=~14 were required, which were calculated by the
program \textsc{EEASiSSS} of \mbox{J.\ Rundgren \cite{Rundgren}}.
This program optimizes a superposition of muffin-tin potentials in a
slab geometry not far from the best-fit structure and also provides
an energy dependent real part of the inner potential $V_\text{0r}$
which varied smoothly by about 8~eV over the large energy range of
the calculations. The damping of the electrons due to inelastic
processes was taken into account by a constant optical potential
$V_\text{0i}$, which was fitted to 5.4~-~\mbox{5.9~eV} for the
different phases grown on iridium and \mbox{5.0~eV} for those on
platinum. As lattice parameters we used
\mbox{$a_\text{Ir}=$~2.7116~{\AA} \cite{iridium}} and
\mbox{$a_\text{Pt}=$~2.7699~{\AA} \cite{platinum}} according to a
data acquisition temperature of 100~K. During the fitting procedure
all geometrical parameters down to the 6$^{th}$ layer were varied
regarding the symmetry of the unit cell ($p2mm$ for 3$\times$1 and
$p4mm$ for 1$ \times$1). Additionally, the vibrational amplitudes of
the surface species were optimized while the vibrations of bulk
atoms were fixed at values of \mbox{$u_\text{Ir}=$ 0.0425~{\AA}} and
$u_\text{Pt}=$ 0.0650~{\AA} calculated according to Ref.\
\onlinecite{VanHove79Buch} from the respective Debye temperatures of
$\Theta_\text{Ir}=$ 420~K and $\Theta_\text{Pt}=$ 236~K.\cite{Debye}
The angular spread within the slightly convergent primary beam used
in the LEED experiment was accounted for by a small off-normal angle
(typically 0.4$^\circ$ for the used LEED optics) and averaging over
4-8 azimuthal directions of incidence.

Though quite a number of structural and non-structural parameters
(up to 33) were fitted in the course of the analyses, the huge
databases collected for each of the phases still ensured very large
redundancy factors in the range $\varrho$ = 12-26 (for details see
Supplemental Material \cite{SupLEED}). The degree of correspondence
between calculated and experimental spectra was quantified by the
Pendry R-factor $R$ \cite{Pendry}, which also allows an estimate of
the statistical errors of each parameter via its variance $var(R)=
R_\text{min}\cdot\sqrt{8\cdot \text{V}_\text{0i}/\Delta E}$. Due to
the low R-factor values achieved for all the LEED analyses, also
very small error margins result which are in the range of 0.01~{\AA}
or below for vertical coordinates of Ir and Co atoms and about
0.03~{\AA} for lateral ones. Since oxygen is a much lighter
scatterer particularly at higher electron energies, the errors are
typically twice as large. The statistical errors for every single
parameter were calculated as usual via single parameter variation
(and thus neglecting possible parameter couplings) and are tabulated
for all LEED analyses in the Supplemental Material \cite{SupLEED}.

\subsection{DFT calculations}
All DFT slab calculations were performed spin-polarized  in analogy
to our former study\cite{Ferstl2016} of these chain phases.
Regarding the spin ordering along the chains, only collinear
ferromagnetic and anti-ferromagnetic configurations have been
considered. We used the Vienna Ab-initio Simulation Package
(VASP)\,\cite{vasp1,vasp2} in the projector augmented wave (PAW)
setup\,\cite{Bloechl1994,Kresse1999}. A plane-wave cutoff of 400~eV
was chosen and exchange correlation effects were treated within the
PBE\,\cite{Perdew1996} approximation adding DFT+U
corrections\,\cite{Dudarev1998} to the Co-$3d$ states with a value
of U-J~=~1.5~eV . The different surface phases were modelled by
asymmetric (3$\times$2) surface slabs with 7~Ir (Pt) layers
separated by at least 14.7~{\AA} vacuum. None of the atomic
positions was kept fixed and all of the atoms where relaxed until
changes in the total energies dropped below 0.1 meV per simulation
cell. A 4$\times$6$\times$1 Monkhorst-Pack type $\bf{k}$-point mesh
was used to sample the Brillouin zone which was increased to
6$\times$9$\times$1 for the calculation of the chain formation
energies and the adsorption energies of additional oxygen atoms. The
formation and adsorption energies are referenced to the free O$_2$
molecule, calculated in its spin-polarized triplet ground state
within a cubic cell of 10~{\AA} side length. Spatial positions
derived from the DFT calculations were scaled by factors of 0.9883
for Ir and 0.9874 for Pt in order to match the calculated bulk
lattice parameters with the experimental ones.

\section{Growth of CoO$_2$ chain phases on Ir(100) and
Pt(100)}
\label{CoO2}

\begin{figure}[htb]
\centering
\includegraphics[width=\columnwidth]{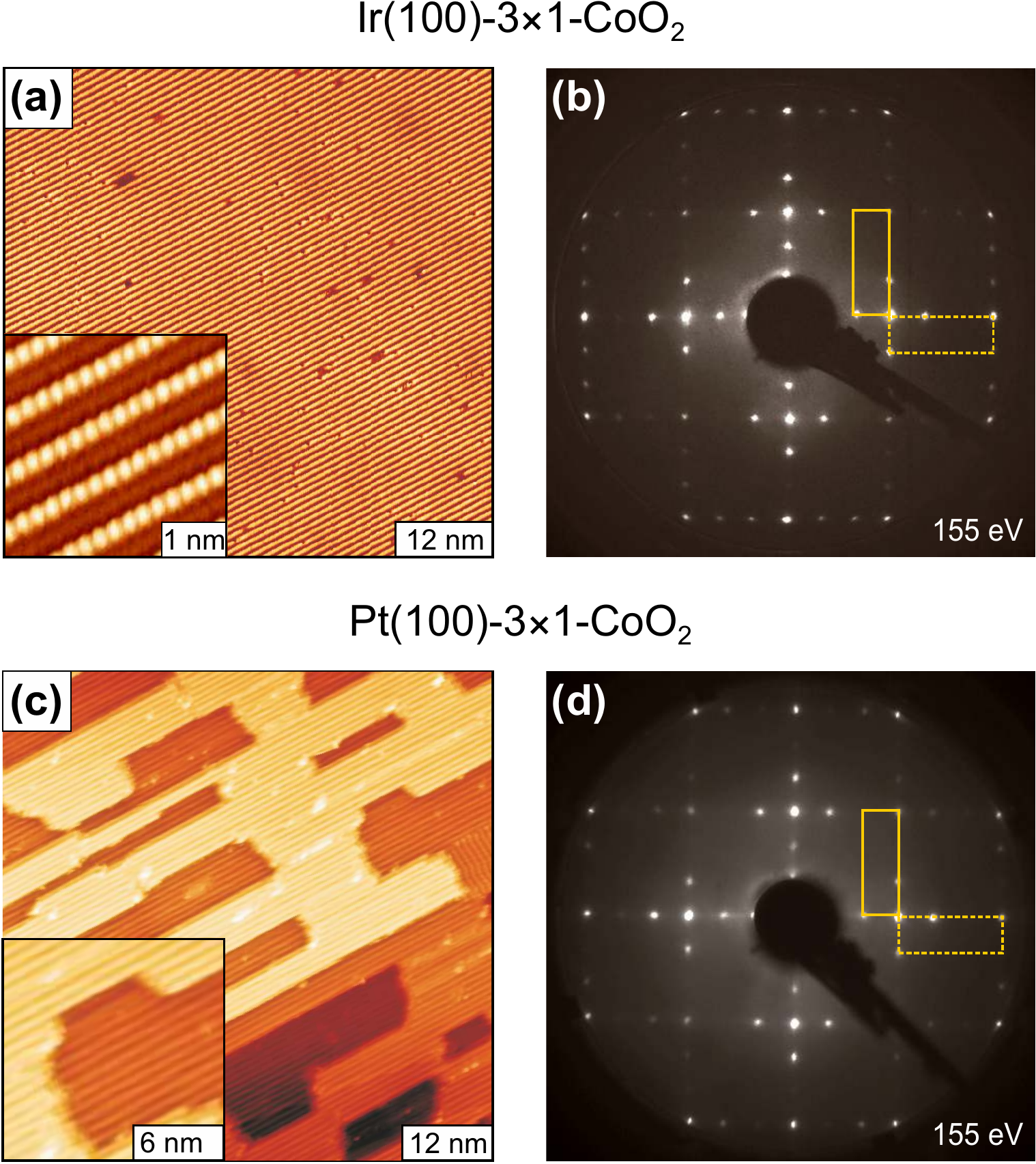}
\caption{(a) STM image and (b) LEED pattern of a homogeneous
3$\times$1-CoO$_2$ phase grown on Ir(100) at 970~K. (c), (d): Same
for a 3$\times$1-CoO$_2$ phase grown on Pt(100) at 730~K. The insets
are close-ups to improve the visibility of the row structures.}
\label{3x1_growth}
\end{figure}

In a preceding publication \cite{Ferstl2016} we have shown that the
oxidation of one third of a monolayer of cobalt (or other transition
metals) on Ir(100) at elevated temperatures leads to the formation
of a perfectly ordered 3$\times$1 phase consisting of
one-dimensional CoO$_2$ chains aligned strictly parallel and
well-separated with a mutual distance of 3$a_\text{Ir}$. The central
monatomic Co wire is shifted in the direction of the chain by half a
lateral unit vector. Below each oxide chain one atomic row of Ir is
missing (\emph{cf.} Fig.~\ref{chain_model}) so that the Co core of
the chains has no direct bond to Ir and interacts with the substrate
only via the oxygen atoms. This reduced substrate interaction of the
chains leads to an at least partially one-dimensional character of
its electronic and in particular of its magnetic properties
\cite{Ferstl2016}. The iridium atoms expelled from the top layer
during the oxidation process (in order to form the missing row
substrate structure) nucleate at temperatures around 650~K into
islands, again with the 3$\times$1 oxide phase on top. Only at
higher temperatures around 950~K the surface diffusion becomes
sufficiently enhanced so that all extra material can be transported
to step edges and plain terraces covered by the oxide phase result
as shown in Fig.~\ref{3x1_growth}(a). Consequently, the
corresponding LEED pattern exhibits very sharp and intense
superstructure spots [Fig.~\ref{3x1_growth}(b)].

The large lateral mass transport required to produce a perfect
3$\times$1 oxide phase explains why there is no qualitative
difference whether the initial cobalt deposition is performed at the
unreconstructed 1$ \times$1 or the reconstructed 5$ \times$1-hex
surface with a 20\% higher atomic density of the topmost layer
\cite{Hove1981,Schmidt2002}. In the latter case the amount of
expelled atoms which have to be removed is just 0.53~ML instead of
0.33~ML for the unreconstructed surface.

The recipe for preparing a homogeneous 3$\times$1-CoO$_2$ phase
developed for Ir(100) can in principle be transferred to the Pt(100)
surface as well. However, on Pt(100) the chains are thermally less
stable and start to decay already at around 770~K, which is 200~K
lower than on Ir(100). This leads to a reduced mass transport of Pt
atoms on the surface even for the highest possible annealing
temperatures. Consequently, Pt adatom islands always remain on the
surface as displayed in Fig.\ \ref{3x1_growth}(c), which restrict
the lateral growth of the oxide chains. The smaller average domain
size of the 3$\times$1 phase on Pt(100) is also expressed in the
LEED pattern [Fig.\ \ref{3x1_growth}(d)] by somewhat broader
superstructure spots compared to the case of Ir(100). Also different
to iridium is the fact that the cool-down in oxygen flux does not
lift the hexagonal reconstruction of the Pt(100) surface. Therefore,
Co deposition has to be performed on the reconstructed surface, so
that about half of the surface is eventually covered by adatom
islands.

\begin{figure}[htb]
\centering
\includegraphics[width=0.95\columnwidth]{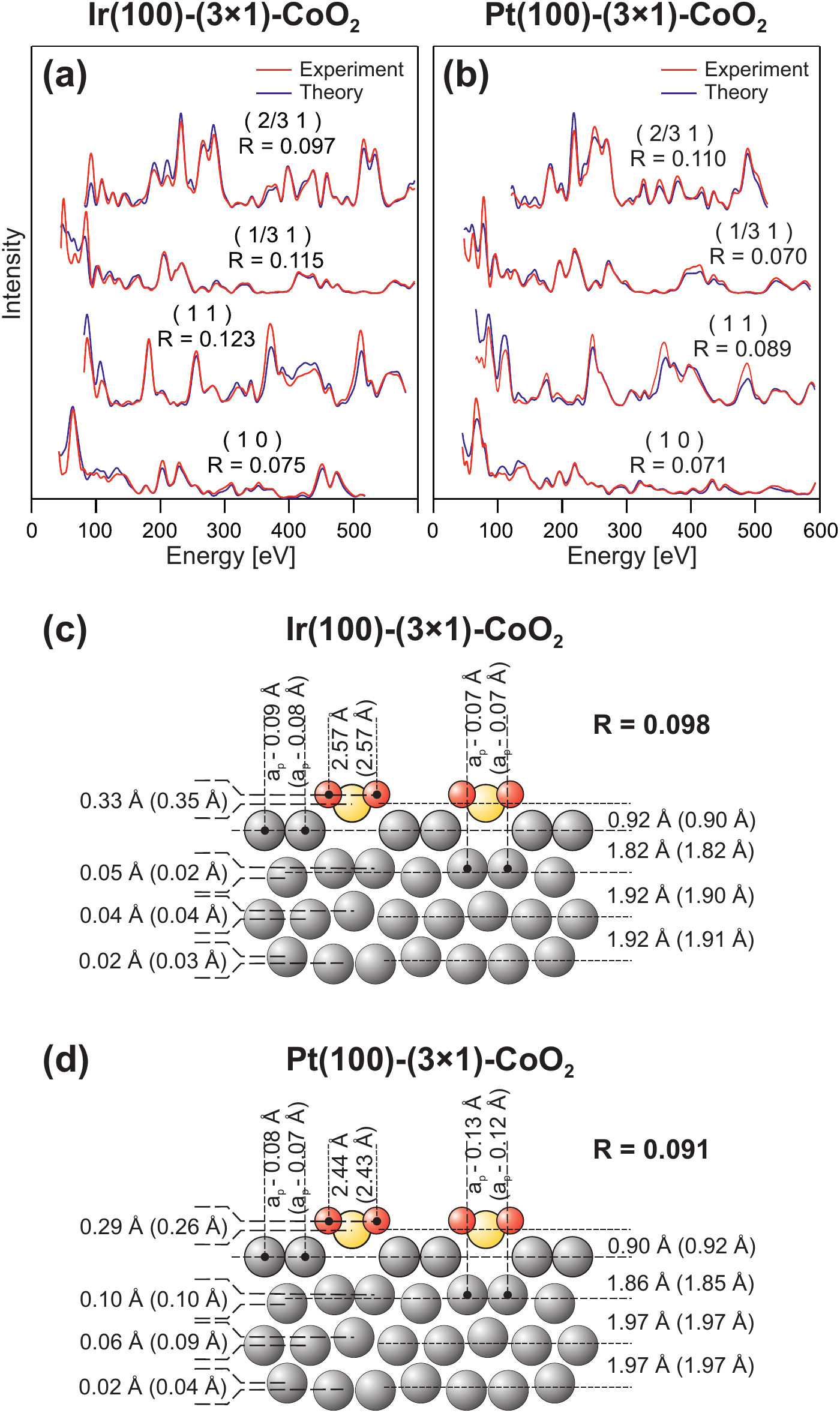}
\caption{(a,\,b): Comparison of experimental and calculated bestfit
$I(E)$-spectra for the 3$\times$1-CoO$_2$ phases grown on Ir(100)
(a) and Pt(100) (b). The Pendry R-factor values given correspond to
the respective beams only. (c,\,d): Corresponding schematic
structural models in side view (vertical distances are strongly
exaggerated) with most relevant geometrical parameters (rounded to
0.01~{\AA}) as derived from the LEED analyses. The corresponding DFT
values are given in brackets. The Pendry R-factors given here refer
to the total data bases used. For a top view see
Fig.~\ref{chain_model}.} \label{ModSpecCoO2}
\end{figure}

A simple comparison of LEED $I(E)$-spectra for both substrates as
shown in the Fig.~\ref{ModSpecCoO2}(a,\,b) reveals strong
similarities in particular for the superstructure spots, which are
especially sensitive to the surface oxide structure. The larger
deviations observed for the integer order spots are mostly due to
the layer distances increased by 0.05~{\AA} for Pt(100) compared to
Ir(100), since the scattering properties of both elements are almost
identical. An independent quantitative LEED analysis for the
Pt(100)-3$\times$1-CoO$_2$ phase -- in analogy to the one performed
earlier for Ir(100) \cite{Ferstl2016} -- showed an excellent
agreement between experimental and calculated $I(E)$-spectra
[Fig.~\ref{ModSpecCoO2}(b)], which is expressed quantitatively by a
Pendry R-factor as low as $R=$~0.091 for the present analysis. (This
also proves that the above mentioned limited size of ordered domains
did not deteriorate the fit quality.) The full set of $I(E)$ spectra
and raw data files can be found in the Supplemental Material
\cite{SupLEED}.

The detailed crystallographic structure of the
Pt(100)-3$\times$1-CoO$_2$ phase is displayed in the
Fig.~\ref{ModSpecCoO2}(d) compared to that for Ir(100) above
[Fig.~\ref{ModSpecCoO2}(c)]. As indicated by the similar
$I(E)$-spectra the structures are nearly identical. In both phases
the oxygen atoms are bound to two cobalt and one substrate atom
each. In case of the Pt substrate the oxygen atoms at both sides of
the Co row have to be closer to each other in order to maintain a
similar Co-O bond length (1.87~{\AA} \emph{vs.} 1.89~{\AA} for Ir)
despite the 0.07~{\AA} larger Co-Co spacing within the oxide chain
(due to the substrate's lateral lattice parameter). Also, the Co row
is lifted up by practically the same amount on both substrates
(0.90~{\AA} \emph{vs.} 0.92~{\AA}) and this essentially planar
binding only via the oxygen atoms leads to exceptionally large
(vertical) vibrational amplitudes for the Co atoms of 0.16~{\AA}
(Ir) and 0.14~{\AA} (Pt) at 100~K. Also the strain-field induced
relaxation pattern within the subsurface layers is qualitatively the
same for both substrates. Only the amplitudes of the local
displacements of substrate atoms from their bulk positions are found
to be almost twice as large for Pt than for Ir atoms, which is in
line with the significantly different elastic moduli of these
elements ($E_\text{Pt}=$ 170~GPa; $E_\text{Ir}=$
528~GPa)\cite{Samsonov2012}. All the structural parameters
determined by the LEED analyses are perfectly reproduced by
independent DFT calculations with a deviation of less than 1.3 pm on
average, i.e.\ well within the limits of error for both methods. A
complete compilation of all bestfit parameter values together with
the corresponding error margins and DFT predictions can be found in
the Supplemental Material \cite{SupLEED}.

\section{Reduction of cobalt oxide chains by H$_2$} \label{Reduction}

\subsection{STM and LEED Appearance} \label{ReductionApp}

\begin{figure}[htb]
\centering
\includegraphics[width=\columnwidth]{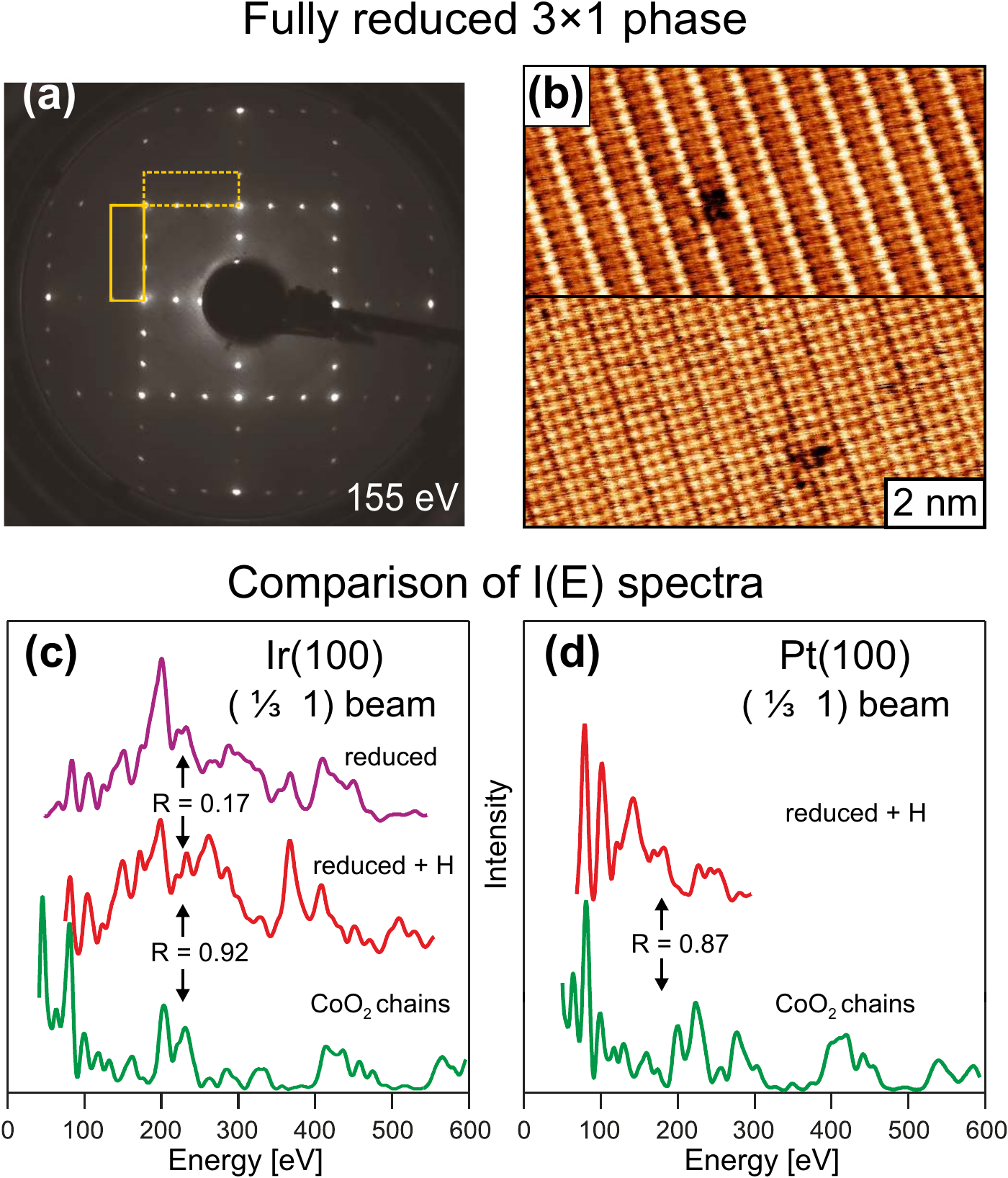}
\caption{(a) LEED and (b) STM appearance of a reduced 3$\times$1
phase on Ir(100) after heating in H$_2$ at 450 K. The two atomically
resolved STM images in (b) were taken under different tip conditions
with (upper part) or without chemical resolution (lower part). (c)
Comparison of $I(E)$-spectra for the ($ \frac{1}{3} \; 1)$ beam
for the reduced phases on Ir(100) with and without coadsorbed
hydrogen and for the regular oxide chain phase. (d) Same for
Pt(100), whereby no clean reduced phase could be stabilized in this
case.} \label{CompPhases}
\end{figure}

Exposing the CoO$_2$ chain phase grown on Ir(100) to hydrogen at
temperatures between 330~K and 570~K does not change the 3$\times$1
periodicity of the LEED pattern (\emph{cf.}
Fig.~\ref{CompPhases}(a)), however, the relative intensities of
diffraction spots are altered as can be seen already by a visual
inspection. These changes become more obvious when comparing
$I(E)$-spectra of superstructure spots as shown for one example in
Fig.~\ref{CompPhases}(c) (green and red curve). R-factor values
close to $R=$ 1 (indicating statistical independence of the data)
prove that the local atomic structure must have changed
substantially within the 3$\times$1 phase. Since hydrogen is a very
weak scatterer the spectral variations cannot be caused by its mere
presence on the surface. Instead, we have to assume that some kind
of reduction process has happened to the oxide chains. In fact, no
desorption of oxygen (or water) can be observed from this new phase
\emph{after} the initial reduction process when the sample is heated
up to a temperature of 1300~K. We thus must have a fully reduced
3$\times$1 phase. However, when heating the sample beyond 370~K some
hydrogen is found to desorb from the sample. This desorption is
accompanied with a quite moderate alteration of the LEED
$I(E)$-spectra [red and purple curves in Fig.~\ref{CompPhases}(c)]
expressed by R-factors in the range of $R=$ 0.2. Such spectral
changes are typical for the impact of hydrogen on the local
positions of substrate atoms being in the range of some picometers.
Obviously, the reduced phase can exist with or without adsorbed
hydrogen.

Further insights into the structure of this reduced 3$\times$1 phase
can be gained from STM. In Fig.~\ref{CompPhases}(b) two different
STM appearances of the reduced phase on Ir(100) are displayed
depending on tip conditions. In the upper part we see one single,
apparently protruding atomic row as for the chain phase [\emph{cf.}
inset in Fig.~\ref{3x1_growth}(a)] though with a considerably
reduced corrugation of only 0.2~{\AA} compared to 0.8~{\AA} before.
The smaller corrugation also allows to image the Ir rows in between,
which turn out to be now in line with the Co rows. This becomes even
more obvious in the lower part of Fig.~\ref{CompPhases}(b), where
chemical resolution is lost. Here, we clearly see triples of atoms
in line with each other, of which the center one must be the Co atom
for symmetry reasons. The triplets appear separated by a somewhat
larger distance, which, however, could also be an electronic
artefact of STM imaging. The interpretation of these images is quite
obvious: Without any oxygen the Co rows cannot reside any more far
above the substrate's missing row structure. Instead, they have to
``fall down'' into the troughs making up new bonds with Ir atoms. In
order to increase the atomic coordination number the Co row as a
whole shifts laterally by half a substrate unit vector to bring the
Co atoms into substitutional sites. In total, the 3$\times$1 phase
converts into an ordered surface alloy with an Ir$_2$Co top layer.
It should be noted that just for steric reasons the postulated shift
of Co rows must be accompanied by the creation of defects like
expelled Co atoms or vacancies within the chains, which cannot be
found in large numbers in the STM images. However, at the
temperatures of the reduction process, where this structural
rearrangement occurs, single Co atoms are already quite mobile at
the surface as known from film growth experiments on Ir(100)
\cite{Giovanardi2008,Heinz2009}. Thus, the defects in the Co rows
can readily heal during the reordering process and so vanish from
the surface.

\begin{figure}[hbt]
\centering
\includegraphics[width=\columnwidth]{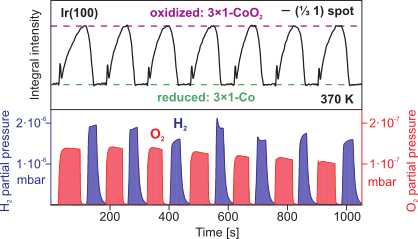}
\caption{Reversible switching between the oxidized
3$\times$1-CoO$_2$ and the reduced 3$\times$1-Co + H chain states on
Ir(100) by variation of the partial pressures of H$_2$ and O$_2$
with the sample held at a constant temperature of 370 K. Top:
Variation of the integral intensity of the ($\frac{1}{3}$ 1)
LEED spot at 65 eV. Bottom: Corresponding partial pressures of
oxygen and hydrogen in the UHV chamber.} \label{Redox}
\end{figure}

From this reduced state the 3$\times$1-CoO$_2$ chain phase can be
recovered by exposing the surface to oxygen at temperatures as low
as 350 K. This can be easily understood since no material transport
is necessary to restore the long range order. The cobalt wires only
have to be shifted and lifted out of the surface. So, the complete
cycle of water formation from molecular oxygen and hydrogen can be
carried out repeatedly at constant sample temperature just by
varying the oxygen and hydrogen partial pressure at the surface as
is demonstrated in Fig.~\ref{Redox}. During this process the actual
state of the surface can be continuously monitored via the intensity
variation of 3$\times$1 superstructure spots. For not too high
partial pressures even the kinetics of the reordering process
including the possible appearance of ordered intermediate phases
(indicated here by the
spikes in the rising slopes of the LEED intensity of Fig.~\ref{Redox}) becomes accessible.\\

\begin{figure}[htb]
\centering
\includegraphics[width=\columnwidth]{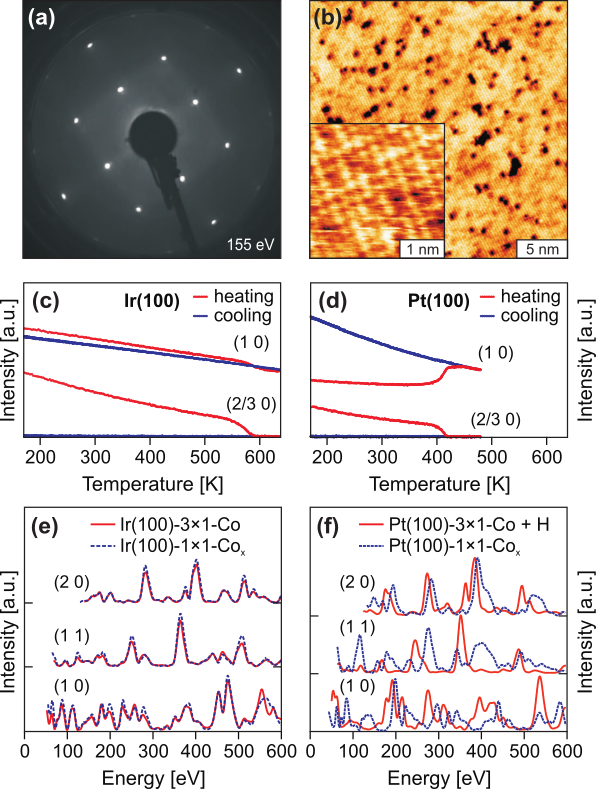}
\caption{(a) 1$\times$1 LEED pattern resulting after heating the
reduced Ir(100)-3$\times$1 phase to 650~K. (b) Corresponding larger
scale STM image with a smaller scale image as inset. (c) Development
of LEED spot intensities taken at 145~eV showing the irreversible
3$\times$1 $\rightarrow$ 1$\times$1 transition with temperature for
Ir(100) and (d) for Pt(100). Starting configurations were the clean,
reduced 3$\times$1 phase in case of Ir(100) and the H-covered one
for Pt(100). (e) Comparison of integer order $I(E)$-spectra for the
(clean or H-covered) reduced 3$\times$1 and clean 1$\times$1 phases
on Ir(100) and (f) on Pt(100).} \label{1x1}
\end{figure}

When the reduced Ir(100)-3$\times$1 phase is heated beyond 570 K or
the reduction process is just performed at these high temperatures,
then the 3$\times$1 superstructure vanishes and a sharp 1$\times$1
LEED pattern can be seen instead [Fig.~\ref{1x1}(a)]. Also in STM
the characteristic stripe-like appearance of the Co chains
[\emph{cf.} Fig.~\ref{CompPhases}(b)] is missing, while atomically
resolved images reveal varying apparent heights of atoms
[Fig.~\ref{1x1}(b)]. The structural transition turns out to be
irreversible since the superstructure spots do not reappear upon
subsequent cooling as is demonstrated by the thermal evolution of
spot intensities in Fig.~\ref{1x1}(c). The integer order spot
intensities, in contrast, are only negligibly affected by the
structural transition, their continuous decrease of intensity with
temperature is simply due to a Debye-Waller damping caused by
enhanced vibrational motions of scatterers. In order to prove that
the insensibility of integer order spot intensities against the
structural transition is not just a matter of the chosen energy we
compare in Fig.~\ref{1x1}(e) the $I(E)$-spectra of 3$\times$1 and
1$\times$1 phases on Ir(100). The perfect coincidence of the curves
is a strong indication that only the long-range order has got lost,
while the elemental distribution within the layers must have been
preserved, a conclusion which will be quantitatively proven by a
LEED analysis discussed in Section \ref{Red_IV}. Thus, after the
annealing step we still have an Ir$_2$Co surface alloy as the
outermost layer, but now it is disordered. Only for quite high
temperatures ($T>$ 1000~K) Co starts to dissolve into the bulk
(driven by entropy). With the finding of a random alloy confined to
the top layer one would expect to see just two different species in
the atomically resolved STM images, which is obviously not the case.
This might be due to a pronounced ligand effect, i.e.\ the apparent
(or even the true) height of the atoms is strongly affected by the
local chemical environment of the respective atom. Finally, it
should be noted that also the disordered surface alloy can be
re-oxidized (and by this re-ordered) into the 3$\times$1-CoO$_2$
phase, though perfect ordering requires somewhat higher temperatures
compared to when starting with the ordered surface alloy. We
observed that even those Co atoms which had dissolved deeply below
the surface upon high temperature annealing will re-segregate and
react during oxygen annealing towards the 3$\times$1-CoO$_2$ phase.

On Pt(100) the reduction of the CoO$_2$ chain phase by H$_2$
proceeds practically identically to Ir(100) as long as the
temperature is low enough to have hydrogen adsorbed on the surface
($T<$ 400~K). Only the ubiquitous presence of adatom islands limits
the degree of achievable order as it was already the case for the
regular chain phase. Again, the intensity spectra of the LEED beams
change dramatically and similar to Ir(100)
[Fig.~\ref{CompPhases}(d)], indicative of an analogous rearrangement
of Co chains upon complete reduction.

The 3$\times$1 alloy phase on Pt(100) also disorders into a
1$\times$1 phase as can be seen in Fig.~\ref{1x1}(d), but already at
a temperature of 420~K, which is 150~K lower than for Ir(100). In
this case the disordering of the surface alloy occurs in parallel to
the desorption of hydrogen. This can be interpreted that on the
Pt(100) surface adsorbed hydrogen is required to stabilize the
3$\times$1 order of the reduced phase. As a consequence no
hydrogen-free 3$\times$1 alloy phase can be prepared on Pt(100).
Another qualitative difference to Ir(100) is the substantial change
of intensities of integer order spots during the order-disorder
transition [\emph{cf.} Fig.~\ref{1x1}(f)]. Although the concomitant
desorption of hydrogen would account for some moderate spectral
variations also of the integer order spots, the observed alterations
of the spectra appear by far too large for that. Instead, a more
substantial atomic rearrangement at the surface has to be expected
as will be proven in the next subsection.

Finally, the re-oxidation by oxygen exposure from both the H-covered
ordered and the clean disordered alloy phase proceeds on Pt(100) as
easily as on Ir(100). So, this system can indeed serve as a
Co/Pt(100) based model catalyst for various kinds of redox
reactions.

\subsection{Crystallographic Analysis of the Reduced Phases} \label{Red_IV}

In order to obtain more insight into the crystallographic structure
of the reduced phases we have performed full-dynamical LEED analyses
for the \emph{clean} alloy phases on Ir(100) and Pt(100). The LEED
analyses of the \emph{H-covered} 3$\times$1 alloy phases including
the associated extended discussion of the H adsorption site(s) are
postponed to a later publication\,\cite{Ferstl_tbp}. Without
revealing all details, the H-covered phases are found to be -- as
expected -- 3$\times$1 ordered Ir$_2$Co or Pt$_2$Co surface alloys
for both Ir(100) and Pt(100).

\begin{figure}[htb]
\centering
\includegraphics[width=0.95\columnwidth]{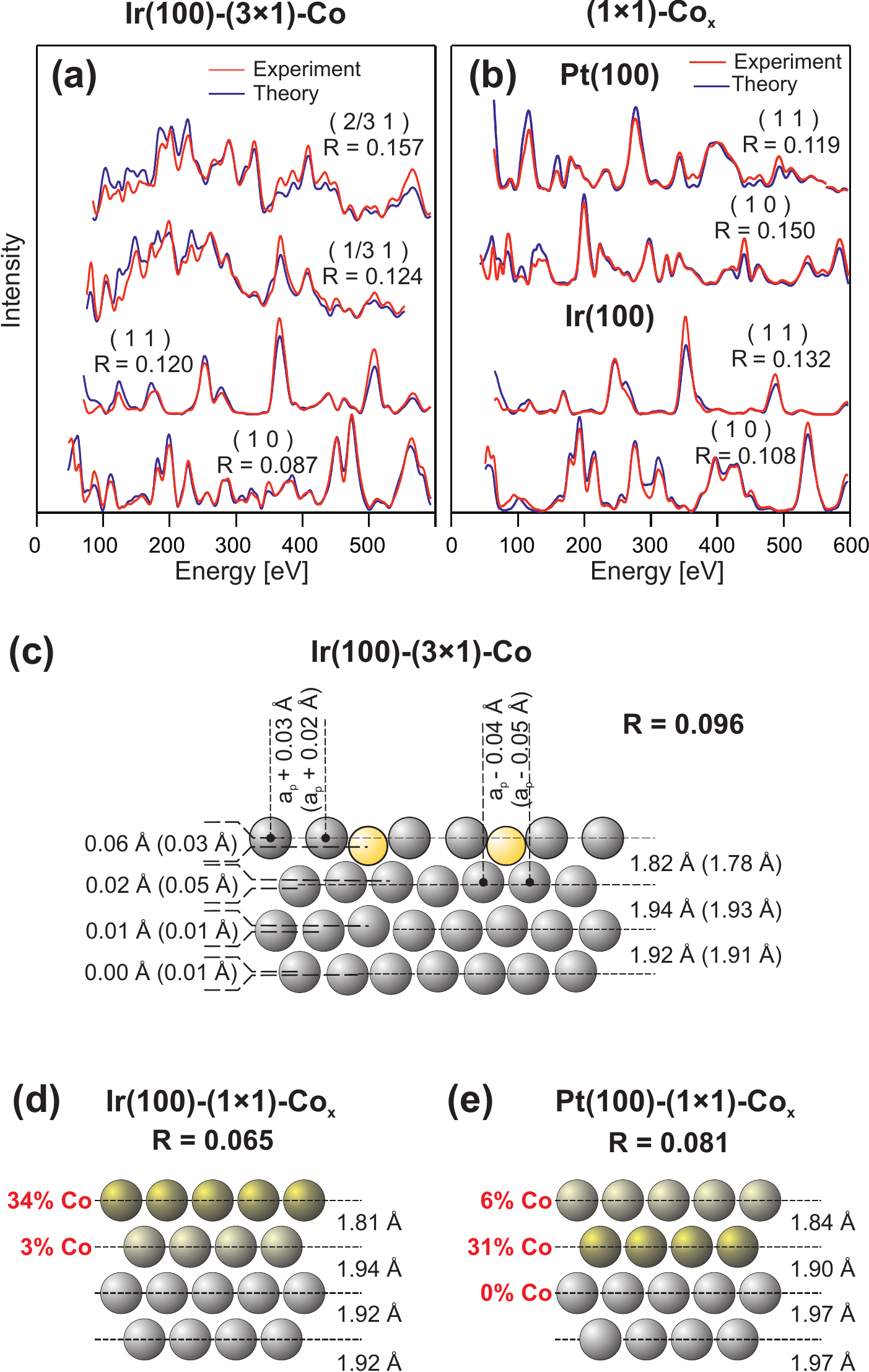}
\caption{(a) Comparison of experimental and bestfit $I(E)$-spectra
for the clean and fully reduced Ir(100)-3$\times$1-Co phase. (b)
Same for the 1$\times$1-Co$_x$ phases on Ir(100) (bottom) and
Pt(100) (top). (c) Schematic structural model for the
Ir(100)-3$\times$1-Co phase with most relevant geometrical
parameters (rounded to 0.01~{\AA}) as derived from the LEED
analysis. The corresponding values predicted by the DFT calculations
are given in brackets. (d) Schematic structural model with
parameters derived from the LEED analysis for the chemically
disordered 1$\times$1-Co$_x$ phases on Ir(100). (e) Same for
Pt(100).} \label{ModSpec_Alloy}
\end{figure}

For the clean 3$\times$1 alloy phase on Ir(100) there is only one
reasonable structural model, for which bestfit parameters have to be
determined: An arrangement of Co rows with threefold mutual distance
assuming substitutional sites within the topmost layer of the
Ir(100) substrate. For this phase we collected $I(E)$-spectra of 35
inequivalent beams in total yielding a huge cumulated data base of
$\Delta E =$ 13.8~keV in total, which safely allowed the
determination of 18 structural parameters down to the 6$^{th}$ layer
as well as of the vibrational amplitudes of Co and Ir atoms within
the top layer (redundancy factor $ \varrho=$~25). Eventually a very
low bestfit Pendry R-factor of $R=$~0.096 was reached by the fit,
quantifying numerically the close correspondence of experimental and
calculated bestfit $I(E)$-spectra displayed in
Fig.~\ref{ModSpec_Alloy}(a). This excellent level of agreement
leaves no room for doubts about the correctness of the underlying
structural model. All details of the LEED analysis including
original $I(E)$-data files can be found in the Supplemental
Materials \cite{SupLEED}.

The structural results obtained for the Ir(100)-3$\times$1-Co phase
are summarized in the ball model of Fig.\ \ref{ModSpec_Alloy}(c). We
find the Co atoms being located 0.06~{\AA} below the level of the
first layer Ir atoms corresponding to the smaller atomic size of Co
compared to Ir. The top layer Ir atoms remain laterally almost at
bulk position (0.01~{\AA} deviation towards the Co atoms) with a
distance to the second layer Ir atoms also close to the distance
found for a pure unreconstructed Ir(100)-1$\times$1
surface\,\cite{Schmidt2002}. This very moderate interaction of Co
with the Ir substrate is also found within the second layer, where
Ir atoms move laterally and vertically by only 0.02~{\AA} towards
the Co atom, when directly attached to it. Consequently, no
significant atomic deviations from bulk position can be found in
deeper layers. As for the analyses of the oxide chain phases the
structural parameters predicted by DFT coincide with the LEED
results similarly well [\emph{cf.} Fig.\ \ref{ModSpec_Alloy}(c)]
with a mean square deviation below 2.1~pm on average. These
deviations are of the order of the error margins for the structural
parameters determined by the LEED analysis, which are tabulated in
the Supplemental Materials \cite{SupLEED}.

Despite the close resemblance of integer order $I(E)$-spectra
between the ordered 3$\times$1-Co and the random 1$ \times$1-Co$_x$
phases we have also performed an independent LEED structure analysis
for the latter phase. In these calculations we treated the random
elemental distribution within the outermost layer(s) by using the
\emph{average t-matrix approximation}
\cite{Gauthier1985,Baudoing1986} and allowed thereby also for
element-specific positions, which is only possible within the
TensorLEED scheme \cite{Heinz1996}. By this we also retrieve error
margins for the layer-resolved concentration of Co atoms. The
analysis achieved a bestfit Pendry R-factor as low as $R=$~0.065 and
hence a very good fitting of the $I(E)$-spectra [\emph{cf.}
Fig.~\ref{ModSpec_Alloy}(b) lower part]. It reveals numerical values
for the Co concentration within the outermost layer of $c_1=$ 34\%
and $c_2=$~3\% in the second layer with errors of about 4\%. This is
a quantitative proof that the irreversible chemical disordering of
the 3$\times$1-Co phase upon annealing is indeed restricted to the
top layer. As expected, the geometrical parameters retrieved by this
analysis coincide very closely with those of the 3$\times$1-Co phase
as seen from a comparison of parameter values given in
Figs.\,\ref{ModSpec_Alloy}(c) and (d). The buckling and pairing
amplitudes of Ir atoms, which will occur only locally within the
random alloy, are reflected by correspondingly enhanced vibrational
amplitudes derived by the fit (see Supplemental Materials
\cite{SupLEED}).

For the random 1$ \times$1 alloy phase on Pt(100) we performed the
very same type of LEED intensity calculation and also achieved an
excellent fit quality [\emph{c.f.} Fig.\ \ref{ModSpec_Alloy}(b)
upper part] expressed by a Pendry R-factor of R = 0.083 in this
case. Surprisingly, in contrast to Ir(100) the fit for this system
reveals that most of the Co atoms have left the topmost Pt layer
where they were located in the ordered H-covered 3$\times$1 phase.
Instead, they now assume substitutional subsurface sites exclusively
within the second Pt layer, i.e. without segregating deeper into the
bulk. The bestfit values for the Co concentration of the outermost
three layers are c$_1$ = 6\%, c$_2$ = 31\%, and c$_3$ = 0\% with
error margins of about 4\%. For the Co atoms within the top layer we
found a pronounced parameter coupling between concentration and
vibrational amplitude leading to unphysical numbers of the fit. We
therefore fixed the amplitude of the Co vibration to 0.10~{\AA},
which is close to the values derived in the LEED analyses for the
other surface alloys. The chemically mixed second layer will of
course impose substantial local geometrical relaxations. In this
sense the layer distances derived by the fit have to be regarded as
averages over various local configurations. Consistent with the
embedding of smaller Co atoms into the second layer both the first
and second (average) layer distance is strongly reduced [\emph{c.f.}
Fig.\ \ref{ModSpec_Alloy}(d)], while deeper layer spacings are
(consistently) rather bulk-like.

A comparable sandwich-like concentration profile, where Co atoms are
(mostly) covered by Pt, has already been found in studies of the
segregation at Pt$_x$Co$_{1-x}$ bulk alloy
surfaces\,\cite{Gauthier1996,Gauthier1998}. Even more, it was
recently found that Co in Pt-Co nanoparticles segregates to the
surface when oxidized but migrates completely back into the
particle's core after reduction in a H$_2$
atmosphere\,\cite{Xin2014}. Eventually, the particles are found to
be coated by a complete Pt monolayer with another pure layer of Co
below, which is very close to our findings and demonstrates the
value of the extended surfaces as a model for nanodisperse
catalysts.

\section{Higher Oxidation State of Cobalt Oxide Chains
achieved by NO$_2$ Dissociation} \label{NO2}

The CoO$_2$ chain phase as described in Sect.~\ref{CoO2} cannot be
oxidized further by exposure to O$_2$ due to an activation barrier
for O$_2$ dissociation unsurmountable under UHV conditions. This can
be circumvented using either atomic oxygen or an oxidant with
smaller dissociation barrier like NO$_2$.

\begin{figure}[htb]
\centering
\includegraphics[width=0.95\columnwidth]{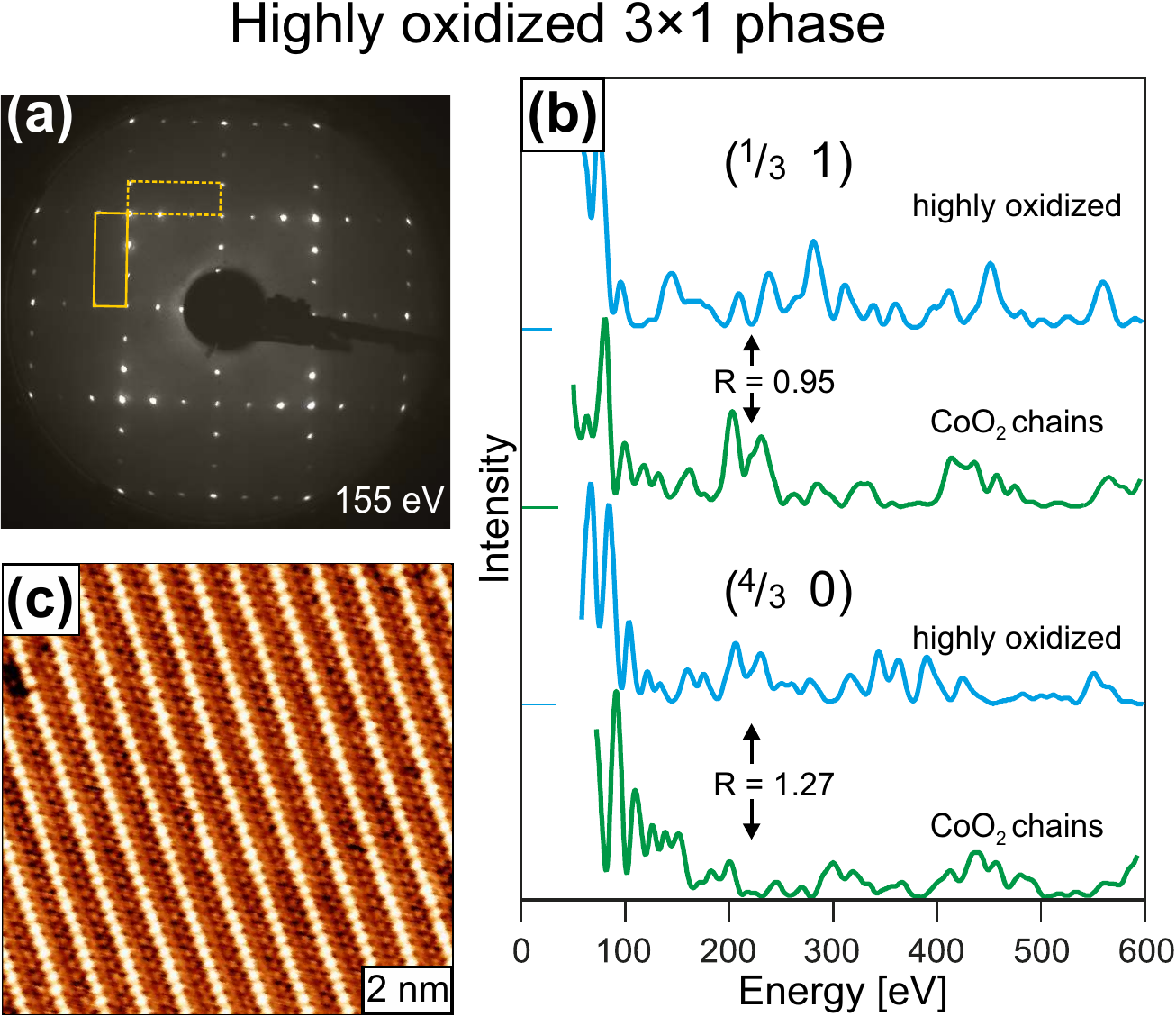}
\caption{(a) LEED and (b) STM appearance of a the highly oxidized
3$\times$1 phase after cool-down in NO$_2$. (c) Comparison of
$I(E)$-spectra for two third order beams of the highly oxidized and
the regular oxide chain phase.} \label{3x1App}
\end{figure}

We cooled the regular CoO$_2$ chain phase from 900~K down to room
temperature in $\sim$10$^{-6}$~mbar NO$_2$. After a further flash to
450~K in order to remove any remaining NO contaminants the LEED
pattern again showed a well-ordered 3$\times$1 superstructure, as
seen in Fig.~\ref{3x1App}(a), but with significantly altered spot
intensities. A comparison of $I(E)$-spectra taken for this new phase
with those for the former CoO$_2$ chain phase [displayed for two
beams in Fig.~\ref{3x1App}(b)] gives R-factor values around $R=$~1,
i.e.\ the spectra are statistically independent. Atomically resolved
STM images from this phase shown in Fig.~\ref{3x1App}(c) look
virtually the same as for the regular oxide chain phase [\emph{cf.}
Fig.~\ref{3x1_growth}(c)]. Even the apparent STM corrugation of the
Co rows -- a quantity which depends quite a lot on the actual tip
and tunneling conditions -- turns out to be at least in the same
range. Careful heating of the new phase to 870~K in UHV leads to
oxygen desorption and by that to a restoration of the regular chain
phase as seen by the respectively changed LEED intensities.
Unfortunately, the exact amount of oxygen released in this process
could not be quantified so far. Nevertheless, this experiment can be
taken as a proof that the new phase contains indeed more oxygen than
the regular CoO$_2$ chain phase. This 3$\times$1 phase can also be
reduced by hydrogen in the same temperature regime as the regular
oxide chain phase. Here, the reduction process directly leads to the
fully reduced 3$\times$1-Co alloy state without passing through the
state of the CoO$_2$ chain phase.

For the Pt(100) substrate we have tried the same preparation recipe
at various sample temperatures and pressures for NO$_2$ exposure.
However, all attempts ended in the regular CoO$_2$ chain phase,
which is obtained already by annealing in O$_2$. The reason for this
failure will be explained in Section \ref{Energetics}. In the
following we thus concentrate on the crystallographic structure of
the oxygen-rich 3$\times$1 phase on Ir(100), which is experimentally
accessible.

\begin{figure}[htb]
\centering
\includegraphics[width=0.85\columnwidth]{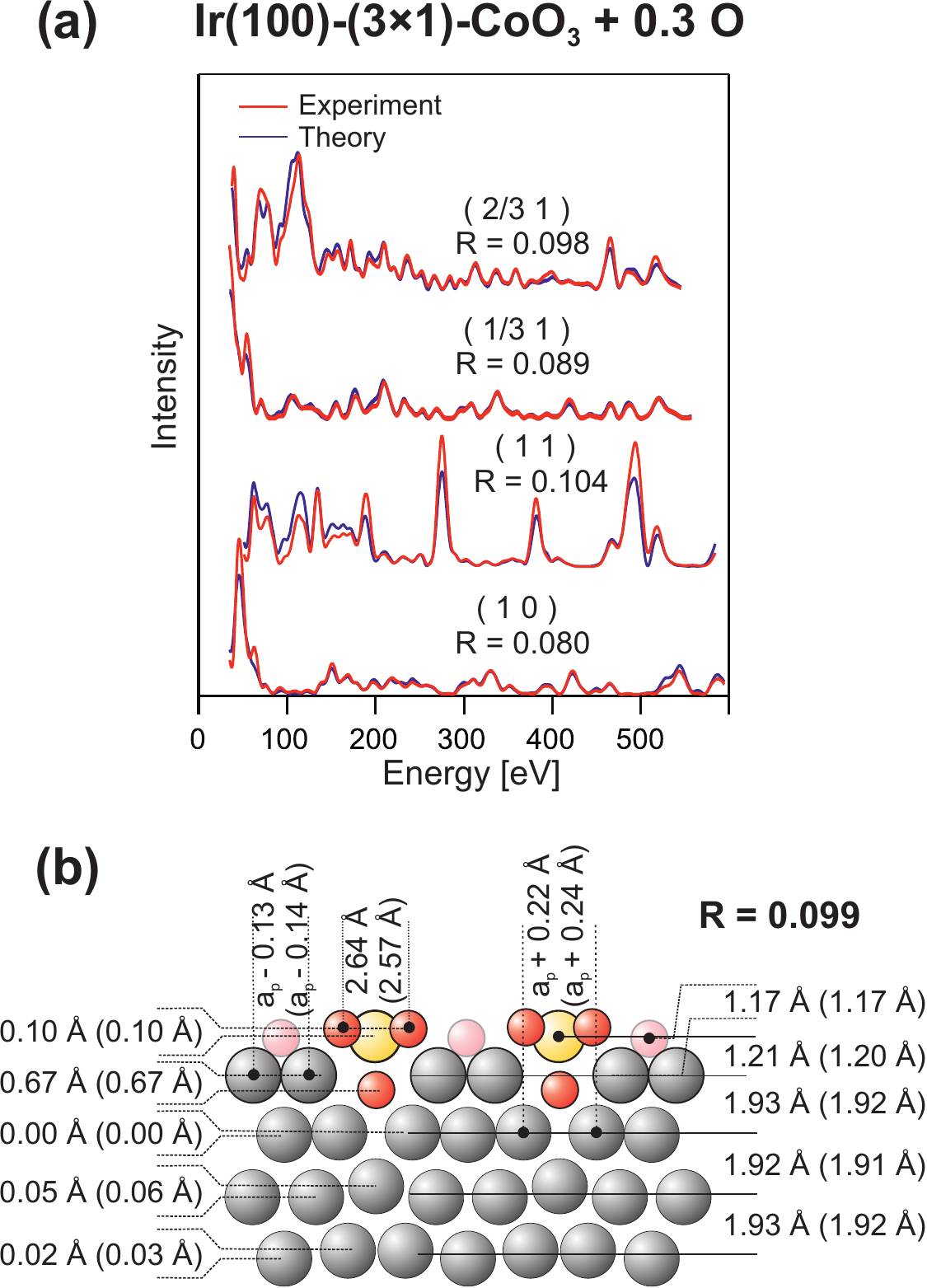}
\caption{(a) Comparison of experimental and bestfit $I(E)$-spectra
for the highly oxidized 3$\times$1-CoO$_3$ phase grown on Ir(100).
(b) Schematic structural model of the CoO$_3$ chain phase (with
coadsorbed oxygen in Ir hollow sites) with most relevant geometrical
parameters (rounded to 0.01~{\AA}) as derived from the LEED
analysis. The corresponding DFT values are given in brackets.}
\label{ModSpecCoO3}
\end{figure}

With this little pre-information about the structure of the new
phase there are a number of qualitatively different structural
models conceivable, which all have to be tested by independent LEED
$I(E) $-fits. In one class of models additional oxygen could simply
occupy bridge or hollow sites on the Ir rows between the normal
CoO$_2$ chains. In a second class of models higher oxidized Co
chains have been tested, i.e.\ with CoO$_3$ chains having the extra
oxygen atom in various positions either above or below the Co atom
or even CoO$_4$ chains, where every Co is enclosed by an oxygen
octahedron. Of course combinations of both classes are also possible
as well as lateral and vertical shifts of the oxide chains relative
to the substrate atoms. It turned out that in a rough search all
tested models (a compilation of the 10 principally different models
under investigation is given in the Supplemental Material
\cite{SupLEED}) produced Pendry R-factors around $R\approx$~0.5 or
even above except one model, which gave $R=$~0.30 and in a further
step of refinement already a value as low as $R=$~0.15. This model
is quite similar to the one for the regular CoO$_2$ chains, but has
an additional oxygen right below each Co atom. We thus have highly
oxidized CoO$_3$ oxide chains at the surface with the Co core still
exposed to the vacuum (truncated octahedron, see
Fig.~\ref{ModSpecCoO3}). In top view such a model exactly resembles
that of the CoO$_2$ chain phase, since the extra oxygen is
completely hidden by the Co atom above. This easily explains the
identical STM appearance of both phases, which would otherwise be
hard to understand at least for most of the tested models.

With this starting point we concentrated on the further refinement
of our model. It turned out that an additional but partial
occupation of Ir hollow sites (30\%) by coadsorbed oxygen brings the
R-factor further down to about $R=$~0.12, while the inclusion of
oxygen at Ir bridge sites or adsorbed NO led to an increase of the
R-factor level. With a final fine-tuning of structural and
non-structural parameters of the analysis we eventually achieved a
bestfit R-factor of $R=$~0.099 derived for a cumulated data base of
about $\Delta E=$~20~keV, which is among the largest ones ever used
for a LEED analysis so far. The again excellent agreement of
experimental and calculated beams is visualized in
Fig.~\ref{ModSpecCoO3}(a) and a ball model in side view is displayed
in Fig.~\ref{ModSpecCoO3}(b) with most of the structural parameters
introduced. In brackets we provide the respective values from a
fully relaxed DFT structure, which was calculated for a 3$\times$2
cell with one oxygen atom in Ir hollow site position, corresponding
to a 50\% partial coverage at this site in slight contrast to the
30$\pm$14\% derived from the LEED analysis. As for the other phases
we find a very close correspondence of parameter values expressed by
a rms deviation of 0.020~{\AA} only. For completion we have also
performed a DFT model calculation for a 3$\times$1-CoO$_3$ phase
without coadsorbed oxygen, which is provided together with all other
information on this structure determination in the Supplemental
Material \cite{SupLEED}. The latter calculation reveals very similar
parameter values as obtained in the case of coadsorbed oxygen except
for the buckling amplitude of the 2$^{nd}$ Ir layer, where a
pronounced 0.06~{\AA} corrugation is predicted. Since such a
buckling is clearly absent in experiment -- the statistical error of
the LEED analysis is below 0.01~{\AA} for this parameter -- we can
take this finding as an independent proof for the existence of the
extra hollow site oxygen.

Turning now to the configuration of the CoO$_3$ chains we see that
the oxygen atoms embedded in the troughs of the substrate's missing
row structure lift the Co atoms located right above by 0.29~{\AA}
almost to the height level of the adjacent oxygen rows. Therefore,
virtually the same O-Co bond lengths result for all oxygen atoms of
the oxide chain (1.88~{\AA} and 1.89~{\AA}) justifying its
assignment as a chain of truncated octahedra. This chain is embedded
in the surface such that also very similar O-Ir bond lengths result
with 1.93~{\AA} and 1.96~{\AA} for the low and high lying oxygen
atoms, respectively. This binding induces severe distortions within
the substrate by pushing all adjacent Ir atoms apart, which leads
via elastic response to statistically significant atomic
displacements down to the 4$^{th}$ layer. In contrast to the CoO$_2$
chains the highly oxidized CoO$_3$ chains are now 3-dimensionally
linked to the substrate, which is also expressed by the now strongly
reduced vibrational amplitude of the central Co atoms (0.09~{\AA}
\emph{vs.} 0.16~{\AA}).

\section{Energetics of the cobalt oxide chain phases} \label{Energetics}

So far the DFT calculations for the various systems were used only
as independent support for the structural models retrieved by the
LEED analyses. In this final section we focus on the energetics of
oxygen adsorption derived from DFT modeling as well as the magnetism
of the monatomic wires in the various phases.

One key quantity for understanding the adsorption behavior is the
binding energy $E_{\text{B}}$ of the adsorbate, here  oxygen, at a
particular site (configuration $\sigma$) with respect to a given
reference state $\sigma_0$ of the surface. We calculate the oxygen
binding energy according to

\begin{equation}
E_{\text{B}}(\sigma) =  -\frac{1}{N_{\text{O}}} [E(\sigma) -
E(\sigma_0) - N_ {\text{O}} \cdot \frac{1}{2} E_{\text{O}_2}]
\label{adsorption-energy}
\end{equation}

for adsorption with respect to  molecular oxygen. $E(\sigma)$ and
$E(\sigma_0)$ are the total slab energies of the adsorbate structure
of interest and of the reference surface configuration.
$N_{\text{O}}$ is the number of oxygen atoms by which the two
configurations differ and $E_{\text{O}_2}$ denotes the DFT energy of
an oxygen molecule in its triplet ground-state. The binding energy
is thus defined positive for  exothermic adsorption.

The initial formation of the mixed Ir$_2$Co top layer is a strongly
activated process, which hardly can be understood on the basis of
mere energetic considerations. We thus take a clean and ordered
3$\times$1-Co alloy phase as reference system and therefore
concentrate in the following on the Ir(100) substrate, where such a
phase is known to be stable. From experiment we also know that the
oxidation of this phase occurs already close to room temperature and
hence all activation barriers should be small, except the one for
O$_2$ dissociation which, however, can be circumvented via NO$_2$
dissociation. In order to determine which adsorption site will be
occupied first by incoming oxygen we calculated its binding energies
in the CoO$_2$ chain phase (which is found in experiment) as well as
for bridge or hollow sites between the two Ir rows for half and full
occupation, i.e. by assuming either all sites or only every second
one in a 3$\times$2 cell (Table~\ref{table1} top). As for the bare
Ir(100)-1$\times$1 surface the Ir bridge site is energetically
preferred over hollow site adsorption and there is also a clear
preference for nearest neighbor site occupation (0.11~eV energy gain
per oxygen) leading to oxygen chain formation \cite{Ferstl2016a}. In
contrast, the Ir hollow site shows an enormous nearest neighbor
repulsion leading to a 0.42~eV reduction of the binding energy, when
filling up all sites. The energetically by far most favourite site,
however, is the CoO$_2$ chain phase, 0.30~eV superior to the Ir
bridge site. This means that oxygen will exclusively be accommodated
within the cobalt oxide chains until the 3$\times$1-CoO$_2$ phase is
completed.

\begin{table}[htb]
\renewcommand{\arraystretch}{1.5}
\centering \vspace{3mm}
\begin{tabular}{lc}
\hline\hline
oxygen adsorption site & $E_\text{B}$ [eV] wrt 3$\times$1-Co\\
\hline
CoO$_2$ chain                     & 2.256 \\

center Ir bridge site              & 1.961 \\

center Ir hollow site              & 1.379 \\

center Ir bridge site half-filled  & 1.849 \\

center Ir hollow site half-filled  & 1.799 \\
\hline
  & $E_\text{B}$ [eV] wrt 3$\times$1-CoO$_2$\\
\hline
CoO$_3$ chain                      & 1.191 \\

center Ir bridge site              & 1.154 \\

center Ir hollow site              & 0.626 \\

CoO$_3$ chain half filled          & 1.014 \\

center Ir bridge site half-filled  & 1.084 \\

center Ir hollow site half-filled  & 1.100 \\
\hline
  & $E_\text{B}$ [eV] wrt 3$\times$1-CoO$_3$\\
\hline

center Ir bridge site             & 1.142\\

center Ir hollow site             & 0.692 \\

center Ir bridge site half-filled & 1.047 \\

center Ir hollow site half-filled & 0.983 \\

\hline\hline
\end{tabular}
\caption{\label{table1} Binding energies $E_\text{B}$ (wrt an O$_2$
molecule) calculated by DFT in a 3$\times$2 cell for oxygen adsorbed
at the specified sites on either the fully reduced 3$\times$1-Co,
(top), the 3$\times$1-CoO$_2$ (middle) or the 3$\times$1-CoO$_3$
(bottom) ferro-magnetic  chain phases on Ir(100).}
\end{table}

In order to test the further course of oxygen adsorption we now
reference the binding energy with respect to an already completed
CoO$_2$ chain phase (center of Table \ref{table1}) and by that we
determine the \emph{effective} binding energy for further oxygen
being adsorbed at the surface. We again compare the energetics of
adsorption sites on the Ir rows with the site right below the Co
atoms, which essentially is the CoO$_3$ phase. For full coverage of
the respective sites, i.e. at a total oxygen coverage of $\theta=$
1, we see that all binding energies for the newly coming oxygen have
roughly halved now. This explains why careful heating of the highly
oxidized phase can reestablish the CoO$_2$ phase, because the extra
oxygen is much weaker bound and thus desorbs at significantly lower
temperatures. Also the energetic difference between the CoO$_3$
chain site and the next favorable Ir bridge position has vastly
diminished to 0.04~eV, which is of the order of $k_\text{B}T$ at
typical reaction temperatures. So there is no more a pronounced
gradient in energy steering the occupation of one particular site.
Moreover, since there is no obvious entrance channel for oxygen into
the subsurface sites below the Co atoms a significant activation
barrier has to be expected for the occupation of these energetically
slightly favoured sites. Conceivable is either an oxygen insertion
at chain defects and subsequent subsurface migration or a site
switch from upper chain oxygen into the subsurface site with
concerted re-filling from Ir bridge sites.

The situation becomes more complicated for smaller coverages like
e.g.\ $\theta=$ 0.83, when only every second of the investigated
sites is occupied. Not unexpected, a half-filled CoO$_3$ chain is
energetically  less favorable, since the equilibrium positions of Co
atoms in local CoO$_2$ and CoO$_3$ configuration are vertically far
apart from each other (about 0.3~{\AA}), which will lead to enormous
stress within such a buckled chain. As a consequence the CoO$_2$ to
CoO$_3$ transformation will preferably proceed for extended chain
pieces as a whole. This points towards an exchange process, which
could act in a zipper-like mechanism, rather than to subsurface
migration from chain defects. More surprising, however, comes the
enormous energy gain for the half-filled hollow site arrangement,
even larger than for the alloy phase despite the largely reduced
absolute values of the binding energies. This together with some
energy loss of the bridge site makes both Ir sites practically
degenerate in energy. However, since both site energies are still
lower than the one for fully covered bridge sites we have to expect
oxygen nucleation towards adatom chains. Regarding the vast increase
of binding energy with decreasing occupation of the hollow sites
(0.47~eV difference between half- and fully filled), it appears at
least conceivable that this trend further continues, so that for
lower coverages a dilute occupation of hollow sites might indeed be
able to compete with bridge site nucleation or even with sites in
the CoO$_3$ chains.

Qualitatively the same scenario is found for the Ir site occupation
in case of a fully developed CoO$_3$ chain phase (lower part of
Table \ref{table1}) and even the resulting binding energies (now
referenced to the completed CoO$_3$ chain phase) are rather close to
the case of the CoO$_2$ phase. This means that the binding energy of
oxygen on Ir sites is hardly affected by the occupation of
subsurface sites nearby. As a consequence the whole oxidation
process of the CoO$_2$ phase can be regarded as a more or less
continuous uptake of oxygen from the gas phase into the Ir sites,
from where it gradually but in a block-wise manner fills up the
subsurface sites. Eventually, this process should end in a complete
CoO$_3$ phase with all central Ir bridge sites occupied as well.
This contrasts with the result of our LEED analysis, which finds
completed CoO$_3$ chains but only 30\% of Ir hollow sites occupied.
An explanation comes by coadsorbed NO molecules evolving in the
NO$_2$ dissociation process. In later stages of the cool-down
process these molecules will not instantaneously desorb from the
surface and thus block adsorption sites for oxygen. With the final
flash in UHV the NO then desorbs leaving an incompletely filled
surface behind. The oxygen atoms remaining at Ir sites do obviously
not nucleate towards bridge-bonded adatom chains but spread out and
occupy hollow sites, which is a corroboration of an enhanced
stabilization of this site for smaller coverages as discussed above.

Regarding the magnetism of the Co wires in the different chain
phases on Ir(100) our calculations find that they are not only
ferromagnetic in the metallic state of the ordered alloy phase, but
also as CoO$_2$ and CoO$_3$ chains though with  decreasing
preference over the anti-ferromagnetic state: $\Delta E =$ 47~meV,
25~meV, and 6~meV for the Co, CoO$_2$, and CoO$_3$ chains.

For the oxidation of Co/Pt(100) we did not perform a similarly
extended analysis as for the Co/Ir(100) system, since both the clean
ordered alloy and the highly oxidized states were not accessible in
experiment. We only calculated the oxygen binding energies
$E_{\text{B}}$ for the pure CoO$_2$ and CoO$_3$ chain phases,
referenced as before to the (here non-existing) ordered
3$\times$1-Co or to the CoO$_2$ phase, respectively, without
considering any oxygen co-adsorption on Pt sites. (The corresponding
structural parameters of these phases are tabulated in the
Supplemental Materials \cite{SupLEED}). The retrieved binding
energies turned out to be significantly smaller for Pt(100) compared
to Ir(100). For the 3$\times$1-CoO$_2$ phase we obtained
$E_\text{B}$ = 2.014~eV (Ir: 2.256~eV) and for the fully oxidized
CoO$_3$ chains a value of $E_\text{B}$ = 0.632~eV (Ir: 1.191~eV)
resulted for the additional oxygen atom. In contrast to the CoO$_x$
chains on Ir(100) the calculations predict an anti-ferromagnetic
ground state for both CoO$_2$ and CoO$_3$ chains on Pt(100).

The lower effective binding energy found for the additional atom in
the CoO$_3$ chain in case of Pt(100) also provides the key to
understanding why we could not prepare this phase under our
experimental conditions. From the binding energies of the different
phases we can calculate the Gibbs free surface energy $\Delta G$
(with respect to the clean 3$\times$1-Co surface alloy) according to

\begin{equation}
\Delta G =  -\frac{N_{\text{O}}}{A} [E_{\text{B}}(\sigma) +
\mu_{\text{O}}] \ \label{free_energy}
\end{equation}

with $N_{\text{O}}$ being the number of oxygen atoms in the units
cell of area A and $\mu_{\text{O}}$ the chemical potential of oxygen
depending on oxygen pressure $p$ and temperature $T$ (Refs.\
\onlinecite{Reuter2001,Rogal2007}):

\begin{equation}
 \mu_{\text{O}} = \mu_{\text{O}}^0 + \frac{1}{2} k_\text{B}T
\ ln\left( \frac{\text{p}}{\text{p}^0}\right) \
\label{chempot}
\end{equation}

The reference state $\mu_{\text{O}}^0$ for the chemical potential
was interpolated from Table 1 in Ref.\ \onlinecite{Reuter2001}, the
reference pressure was taken as $p^0=$ 1 bar and $k_\text{B}$
denotes the Boltzmann constant.

\begin{figure}[htb]
\vspace{5mm} \centering
\includegraphics[width=\columnwidth]{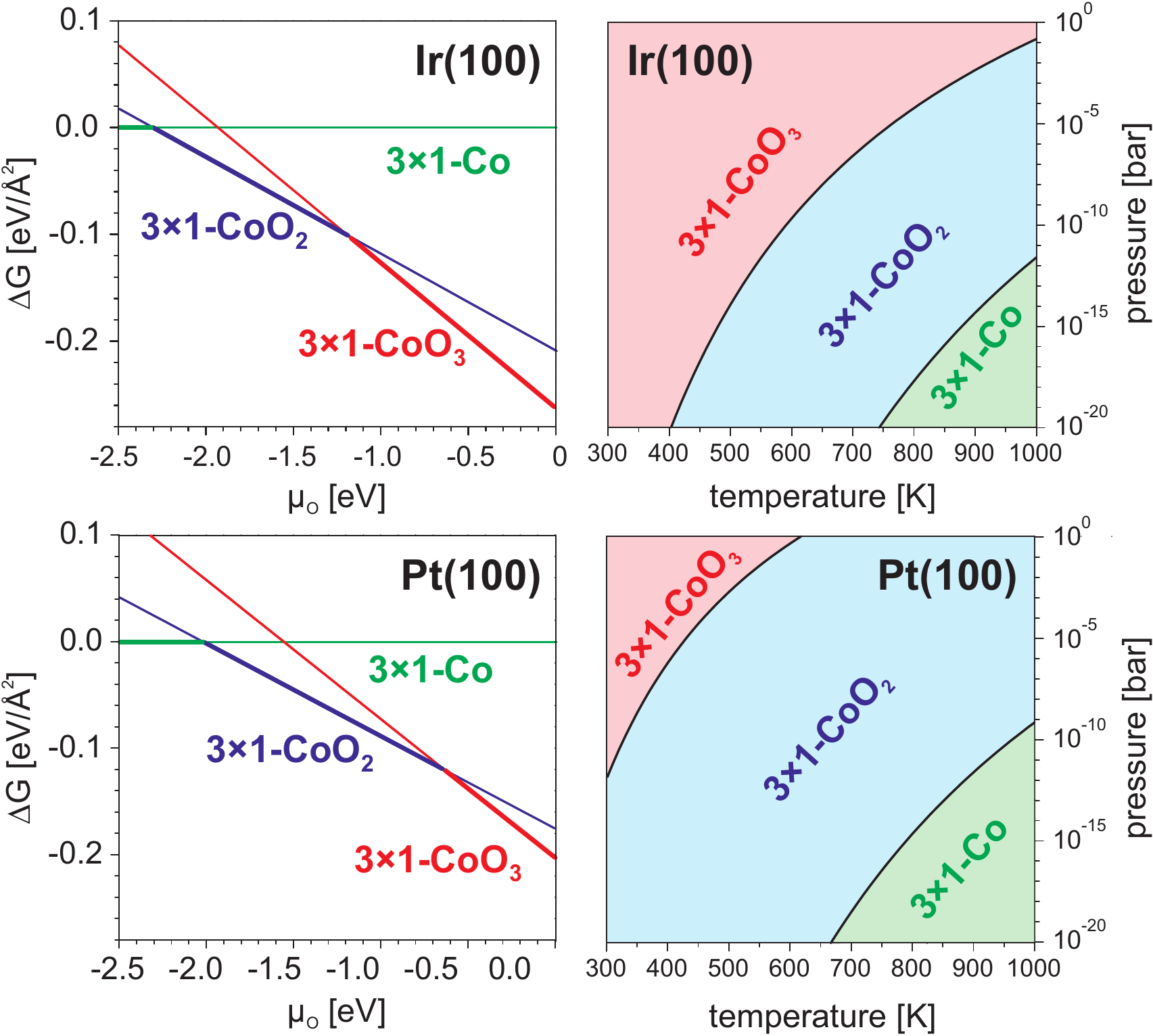}
\caption{Left: Gibbs free surface energy \emph{vs.} oxygen chemical
potential of the chain phases on the Ir(100) (top) and the Pt(100)
surface (bottom) disregarding coadsorbed oxygen. The energetically
lowest state is marked by bold lines. Right: Corresponding $p$-$T$
phase diagrams calculated according to Eq.\ \ref{chempot}.}
\label{Stability}
\end{figure}

Comparing the phase stability for both substrates (Fig.\
\ref{Stability}, left side) we see that for the Pt(100) surface a
higher chemical potential of oxygen than on Ir(100) is required to
stabilize both the CoO$_2$ and in particular the CoO$_3$ phase
explaining the experimentally observed differences in the phase
formation. Transferring the respective stability ranges into the
$p$-$T$ diagrams displayed at the right side of Fig.\
\ref{Stability} we see that on Ir(100) the CoO$_3$ chains are
predicted to be stable under UHV conditions even for rather high
temperatures up to about 550~K. It also shows a wide temperature
window above, where the CoO$_2$ phase is stable in accordance with
the experimental finding that it can be re-established simply by
heating from the highly oxidized state. On the Pt(100) surface, in
contrast, oxygen pressures in the millibar regime would be necessary
to stabilize the CoO$_3$ phase at temperatures above 450~K, which
are needed to surmount activation barriers in the formation process.
So, even in case that such CoO$_3$ chains could be formed in the
NO$_2$ atmosphere of our experiments, they would immediately decay
during pump-down of the reaction gas. However, at more ambient
pressures as e.g. under the conditions of realistic catalytic
processes the highly oxidized CoO$_3$ chains are predicted to occur
on Pt(100) as well. The CoO$_2$ phase on the other hand is stable
under UHV also on Pt(100), though only up to less high temperatures
than on Ir(100) again in correspondence to the different stability
regimes found in experiment. \vspace{5mm}

\section{Conclusions}

In conclusion we demonstrated the formation of strictly periodic,
high-density Co nanostructures with variable numbers of oxygen atoms
bound to Co supported on Ir(100) or Pt(100). Starting from monatomic
CoO$_2$ chains in a 3$\times$1 periodicity that were investigated
previously on Ir(100) and shown here also to exist on Pt(100), we
obtained a purely metallic alloy after reduction with hydrogen. In
the case of Ir(100) the alloy is well-ordered with the same
periodicity and stable up to almost 600~K while on Pt(100) the alloy
is ordered only in the presence of hydrogen and disorders above
420~K. The transition from CoO$_2$ to Co chains is reversible in an
oxygen atmosphere, the well-ordered CoO$_2$ chains reappear with
high structural order from 350~K onwards. Alternatively, the CoO$_2$
chains can be converted on Ir(100) to CoO$_3$ chains under UHV
conditions using NO$_2$ as an oxidizing agent.

All structures appearing were characterized in a crystallographic
sense with LEED intensity analyses to high accuracy with Pendry
R-factors below R = 0.10 and errors in the structural parameters of
a few picometers. The structures were independently confirmed by DFT
calculations with perfect agreement between parameters also in the
pm range. To gain further insight into the energetics of the
formation of these structures, our DFT calculations show that the
CoO$_2$ and CoO$_3$ oxide chains are indeed preferred over
configurations where additional oxygen is adsorbed on Pt or Ir sites
of the respective structure with lower oxygen content. According to
our DFT calculations the CoO$_3$ phase is also stable on Pt(100)
however only at pressures in the millibar regime at reaction
temperatures. The calculations also reveal interesting magnetic
properties of the chain structures as function of substrate and
oxygen content. While the clean metal alloy phases are predicted to
be ferromagnetically ordered on Ir(100) and Pt(100), the CoO$_2$ and
CoO$_3$ phases are ferromagnetic on Ir and anti-ferromagnetic on Pt.
With this work we therefore provide well characterized model systems
for further experiments in heterogeneous catalysis, in
low-dimensional magnetism, and generally for experiments with
low-dimensional metal/metal-oxide nanostructures.

\vspace{5mm}

This work was supported  by the Austrian Science Fund (Project No.
45, Functional Oxide Surfaces and Interfaces, FOXSI) and the
Deutsche Forschungsgemeinschaft (Research Unit FOR 1878
``\emph{funCOS}'' and the DACH project ``\emph{COMCAT}''). We also
thank the Vienna Scientific Cluster (VSC) for generous supply of CPU
time.

\end{document}